\documentclass[epj]{svjour}
\usepackage{graphics}
\usepackage{graphicx}
\usepackage{rotating}
\usepackage{longtable}
\usepackage{lscape}
\usepackage{color}

\begin{document}

\title{EUV magnetic-dipole lines from highly-charged high-Z ions with an open
$3d$ shell}

\author{D.~Osin \and J.D.~Gillaspy \and J.~Reader \and
Yu.~Ralchenko\thanks{\email{yuri.ralchenko@nist.gov}}}

\institute{National Institute of Standards and Technology, Gaithersburg, MD 20899-8422, USA
}

\date{Received: date / Revised version: date}

\abstract
{
The electron beam ion trap (EBIT) at the National Institute of Standards and
Technology was used to produce highly-charged ions of hafnium, tantalum and
gold with an open $3d$ shell. The extreme-ultraviolet (EUV) spectra from these
ions were recorded with a flat-field grazing-incidence spectrometer in the
wavelength range of 4.5 nm to 25 nm. A total of 133 new spectral lines,
primarily due to magnetic-dipole transitions within the ground-state $3d^n$ 
configurations of the
Co-like to K-like ions, were identified by
comparing energy-dependent experimental spectra with a detailed
collisional-radiative modeling of the EBIT plasma.
}
\maketitle

\section{Introduction}
\label{sec:intro}

Accurate knowledge of structure and spectra of highly-charged ions (HCI) is of
great importance for atomic physics, astrophysics, and other fields of
research \cite{Beyer_Shevelko_Intro_HCI_2003,Gillaspy_HCI_2001}. During the
last decade there was a surge in spectroscopic studies of HCI of heavy
elements for fusion applications \cite{9823EL}. This effort is primarily
motivated by the proposed use of tungsten as a plasma-facing material in the
divertor region of the international reactor ITER \cite{Hawryluk09}. Atoms of
tungsten will be sputtered from the divertor plates and are expected to deeply
penetrate the plasma core. The core temperatures on the order of 10-20 keV are
not sufficient to completely ionize tungsten, and therefore the partially
ionized atoms will strongly emit in the x-ray and extreme ultraviolet (EUV)
ranges of spectra. Although this will result in undesirable radiative power
losses, on the positive side, the measured radiation can be reliably used to
diagnose such plasma properties as, for example, temperature and density. This
potential application stimulated an extensive analysis of W spectra (see,
e.g., \cite{16569EL} and references therein), which resulted in identification
of a large number of new spectral lines and in the development of new
techniques for diagnostics of very hot plasmas \cite{12366EL,16374EL,12263EL}.

The spectroscopic properties of other high-Z elements ($Z \sim 70-80$) are
also a subject of active research. The interest in high-Z emission is not
limited to the isoelectronic analysis of spectra with a goal of a better
understanding of W emission. Since gold is one of the primary materials for
hohlraums used  in the inertial confinement fusion experiments \cite{NIF2012},
its highly-charged ions were studied in electron beam ion traps (EBITs)
\cite{12366EL,16347EL,11929EL,11933EL}, laser-produced plasmas
\cite{Reader_HC_Cu_Zn_1980,Reader_HC_Ba_W_1981,6846EL,10546EL,9170EL}, and
tokamaks \cite{11118EL,9820EL,Putterich_PhD_HCI_Tokamaks_2006,12530EL}. Also,
the alloys containing tantalum are considered to be another potential
candidate for a plasma-facing material in tokamaks, and therefore the physical
properties of Ta under the influence of hot plasmas are being examined too
\cite{Hirai_etal_TEXTOR_Ta_2003}. Our group has recently reported analysis of
EUV spectra of Hf, Ta and Au from the EBIT at the National Institute of
Standards and Technology (NIST) \cite{Draganic_etal_Hf_Ta_Au_2011}, where more
than 100 new spectral lines from 35- to 52-times ionized atoms were
identified. Those spectra are mainly due to the intrashell n=4--n=4
transitions in the ions with an open $4s$ or $4p$ shells.

Here we present measurements and identifications of EUV spectra from even
higher (ion charge $z$=44-60) ions of Hf, Ta and Au with an open $3d$ shell in
the ground configuration. The analogous spectra from tungsten ions were
studied recently in Ref.~\cite{16374EL}, where we found that practically all
spectral lines between 10 nm and 20 nm are due to the forbidden
magnetic-dipole (M1) transitions within the ground state configurations
$3d^n$. It was also found that many of the corresponding line ratios show high
sensitivity to electron density in the range typical for fusion plasmas and
thus can be used for diagnostics. In this work we continue application of the
detailed collisional-radiative modeling to the analysis of spectral line
intensities in EBIT plasmas and use it to identify the measured lines.

\section{Experimental setup}

The measurements of the EUV spectra of 3d$^n$ ions of Hf, Ta and Au were
carried out in the NIST EBIT, which produces an electron beam with a beam
diameter of about $60~\mu$m and a current density of about 3500~A~cm$^{-2}$.
The electron energy is controlled by applying a voltage to the central drift
tube, and can be precisely varied in a range between 100~eV and 30~keV.  The
uncertainty in the electron beam energy is on the order of $\pm$ 50~eV, which
is mainly caused by the space charge of the beam \cite{7624EL}. The design and
operation principles of the EBIT are described in more detail elsewhere
\cite{Gillaspy_EBIT_1997}. In these  measurements, the electron beam energy
was varied in a relatively narrow region between 4.0~keV and 6.5~keV for Hf
and Ta, and $5.0$ keV to $7.5$~keV for Au (Table \ref{TabEb}). This set of
energies allowed us to produce spectra of all $3d^n$ ions of Hf, Ta and Au.
The electron beam current was kept constant at 150~mA for all energies, the
central drift tube potential was 220~V, and the ion loading duty cycle was 11
seconds throughout the entire set of measurements.

The metallic ions were introduced into the trap using the metal vapor vacuum
arc ion source (MEVVA) \cite{Holland_MEVVA_2005}. Neutral gases, used mainly
for wavelength calibration, were loaded into the trap through the gas injection
system \cite{12184EL}.

The spectral window observed in the course of these measurements extended from
about 8~nm to 26~nm for Hf and Ta, and from 4.5~nm to 19.5~nm for Au. The
spectra were recorded with a flat-field grazing-incidence EUV spectrometer,
which is extensively described in \cite{Blagojevic_2005}. The emitted
radiation was collected from the 2-cm central region of the EBIT by a
gold-coated spherical mirror, which focused the light onto the spectrometer
entrance slit. The EUV spectrometer is equipped with a grazing incidence,
aberration-corrected, variable-line-spacing grating, which has
1200~lines~mm$^{-1}$ in the center, and which produces a 2-D
spectrally-dispersed image in a plane rather than on the Rowland circle
\cite{Harada_FFgrating_1980}, \cite{Kita_FFgrating_1983}. The vertical slit
was kept wide-open in order to preserve a constant resolving power of about
400, as observed in  \cite{Blagojevic_2005}, and in order to collect as much
light as possible. The wide-open entrance slit, on the other hand, permits
observation of the EBIT ion cloud shifts, which contribute to systematic error
in the wavelength measurements. The spectra were recorded with a windowless
nitrogen-cooled back-illuminated charge-coupled-device (CCD) camera, which has
a spectroscopic type chip with $1320\times400$ pixels of
$20~\mu$m$\times20~\mu$m each. The total spectrum at every beam energy is
combined from 20 one-minute spectra. This allows for easy removal of cosmic
ray traces. In order to increase the signal-to-noise ratio, each individual
spectral image is integrated over pixel columns aligned parallel to the image
of the trapped ion cloud.

The measured spectra of Hf and Ta ions, observed in the region from 8~nm to
26~nm, were calibrated with known lines \cite{NIST_ASD} of N$^{3+}$-N$^{6+}$,
Kr$^{17+}$-Kr$^{33+}$, O$^{4+}$-O$^{5+}$, Fe$^{16+}$-Fe$^{23+}$, and
Xe$^{39+}$-Xe$^{43+}$. For the Au spectra, calibration lines of
Ar$^{7+}$-Ar$^{8+}$ and Ne$^{5+}$-Ne$^{7+}$ were used in addition to the O,
Fe, and Xe lines. It should be mentioned that some calibration lines of highly
ionized Xe were identified in our previous EBIT run
\cite{Draganic_etal_Hf_Ta_Au_2011} (see also \cite{XeTBP}). Both the MEVVA and
the gas injector were utilized to introduce the calibration elements into the
EBIT trap. The calibration spectra were measured in a relatively broad range
of beam energies from 1250~eV to 9300~eV in order to cover the entire spectral
range. 

The calibration lines were fitted with statistically-weighted Gaussian line
profiles.  The calibration curve was obtained by the weighted fit of the line
center positions determined in CCD pixels to the known wavelengths using a
fourth-order polynomial. The weighting in the fit contained contributions from
the Gaussian fit uncertainties, the uncertainty in the assigned calibration
wavelengths, and the estimated systematic measurement uncertainty (a constant
value in CCD pixels, which was converted into wavelength using the known
dispersion of the spectrometer \cite{Blagojevic_2005}).  The quadrature sum of
the three uncertainty components yielded the final uncertainty of the
wavelength. The statistical uncertainties in line positions were typically
less than 0.001 nm. Multiple wavelength values observed at different beam
energies were weight averaged. The total error in the final wavelength was
taken to be the quadrature sum of the total uncertainty from the calibration
curve and the statistical uncertainty from the weighted average. The final
uncertainty of the Hf, Ta and Au spectral lines was mainly on the order of
0.003 nm and somewhat higher for blended lines.

Figures \ref{Fig1}-\ref{Fig3} show the evolution of the measured spectra of
Hf, Ta and Au with the beam energy. The signal counts in figures are the CCD
analog-to-digital units (ADU). The high signal-to-noise ratio obtained in the
spectra was important for identification of blended and weak spectral lines.
The spectra contain a number of lines from impurity elements. The NIST EBIT is
routinely used for various other studies including Xe deposition on surfaces,
and therefore xenon lines are almost always seen in our spectra. For instance,
several lines from the Xe ions with ion charge $z\approx$ 40 are observed
between 15 nm and 17 nm. Also, several lines from highly-charged O and Ar can
be identified as well. In some cases the impurity lines blend with the lines
from Hf, Ta and Au (e.g., Ar XV line at 22.115 nm with 22.091 nm line from
Cr-like Hf). However, their contribution can be reliably isolated using the
energy dependence of spectra and comparison with the simulated spectra which
do not include impurities. Some of the impurity lines are indicated in
Fig.~\ref{Fig1}.

% -------------------------------------------
\section{Collisional-radiative modeling and line identifications} 

As with our previous studies \cite{12366EL,16374EL,9000EL,12286EL,14870EL},
the spectral analysis and line identifications are based on detailed
collisional-radiative (CR) modeling. In \cite{16374EL} we comprehensively
described the basic principles of CR simulations for W ions with an open $3d$
shell, and the present simulations follow that approach. Briefly, we start
with calculation of all required atomic data (energies, radiative
probabilities and electron-impact cross sections) using the Flexible Atomic
Code (FAC) \cite{FAC}. The CR model includes singly-excited states up to at
least $n$=5 (up to $n$=8 for ions near closed shells) as well as
doubly-excited states with $n$=3. In order to reduce the number of states
included, the $n \geq 4$ levels are combined into ``superterms," which were
introduced in \cite{16374EL}; this results in more manageable matrix sizes
without loss of accuracy. The energies of the singly- and doubly-excited
states within $n$=3, which are the most important for the present study, are
improved by performing additional extensive calculations which include all
possible configurations of the $n=3$ complex as well as the configurations
described above. The calculated atomic data is then used as input to the
non-Maxwellian CR code NOMAD \cite{NOMAD}, which calculates ionization
distributions (also taking into account charge exchange between ions and
neutrals in the EBIT), level populations, and spectral line intensities.
Finally, the theoretical spectrum is convolved with the spectrometer
efficiency curve and compared with the measured spectra. This procedure allows
us to identify the spectral lines using not only their wavelengths but also
their line intensities.

The calculated ionization potentials for the Ni-like to Cl-like ions of Hf, Ta
and Au are presented in Table~\ref{TabIP}. Also the Dirac-Fock results of
Ref.~\cite{8672EL} are given for comparison. The two sets of ionization
energies are seen to agree within 0.4\% or less. One can see then that the
electron beam energies of Table~\ref{TabEb} are sufficient to produce the
required ionization stages with $3d^n$ ground configurations.

The experimental spectra in Figs.~\ref{Fig1}--\ref{Fig3} cover a quite large
range of wavelengths. This is achieved with a moderate spectral resolution
$\lambda / \delta \lambda \sim$ 400 that results in a number of blended lines.
In such cases, the corresponding wavelengths were determined from those
spectra where the abundance of one of the ions is small. This procedure is
exemplified by the line at $\sim$ 17.3~nm which was found to be very strong
between 4350 eV and 5035 eV (Fig.~\ref{Fig1}). Closer inspection (see
Fig.~\ref{Fig1b}) shows that at lower values of the beam energy $E_b$, the
main contribution comes from the 17.263~nm line in the Mn-like ion, while at
higher energies only the 17.300~nm line in the Cr-like ion survives.
Accordingly, the wavelengths of these two lines were derived from the spectra
measured at the lowest or highest energies of the beam. Figure~\ref{Fig1} also
shows a strong line from Fe-like Hf$^{47+}$ at 17.424~nm that is blended with
an impurity line at 17.438~nm. In order to analyze possible impurities, we
routinely measured EBIT spectra without the metal ion injection. Such
background spectra at approximately 5000 eV show a clear presence of an
impurity at 17.438~nm; the wavelength of this line was derived from a Gaussian
fit, and its contribution to the wide line profile with Hf injection was
accounted for in a two-Gaussian fit.

Tables \ref{TabHf}-\ref{TabAu} present the measured wavelengths, line
identifications, theoretical wavelengths and transition probabilities for 135
identified spectral lines, 133 of which are new. The identifications
calculated by FAC are given in jj-coupling, and the notations used are
described in \cite{16374EL}. For such highly-charged ions this type of
coupling is the most appropriate. In the tables, $l_{\pm}$ denotes an electron
state with the total angular momentum $j$=$l\pm1/2$, so that $d_+$ corresponds
to $j$=5/2 and so on. Also, the wavelength uncertainties in the tables are
given in the units of the last significant digit, so that 16.281(4) means
16.281$\pm$0.004.

\subsection{Hf spectra}

Table \ref{TabHf} presents 43 new spectral lines from Ni-like Hf$^{44+}$ to
Ar-like Hf$^{54+}$. Except for the four M1 lines within the excited
configurations $3p^53d^2$ of K-like and $3p^53d$ of Ar-like ions, almost all
other lines correspond to the M1 transitions within the  ground state
configurations $3d^n$, $n$=1--9. The only non-M1 transition identified in the
Hf spectra is the electric-quadrupole (E2) line at 16.149~nm in Ca-like
Hf$^{52+}$, for which the M1 transition is not allowed by the $|\Delta J| \leq
1$ selection rule. The lowest-energy spectra in Fig.~\ref{Fig1} also contain
several lines from Cu-like and Ni-like ions which have been analyzed in our
previous work \cite{Draganic_etal_Hf_Ta_Au_2011}. In the present work the M1
transition within the excited configuration $3d^94s$ of Ni-like Hf$^{44+}$ was
identified at 21.944$\pm$0.003~nm; the recent result of 21.9377~nm from
the relativistic many-body perturbation theory (RMBPT) \cite{Safronova_2007}
thus reasonably agrees with our value. The measured wavelength for the strong
resonance line $4s$--$4p$ in the Cu-like ion is 13.372$\pm$0.003~nm which
excellently agrees both with our previous result of 13.373$\pm$0.003~nm
\cite{Draganic_etal_Hf_Ta_Au_2011} and with the older tokamak measurement of
13.375$\pm$0.005~nm \cite{Putterich_PhD_HCI_Tokamaks_2006}. 

The only two lines for which there exist theoretical wavelength calculations
are the $3d^9_{5/2}$--$3d^9_{3/2}$ transition in the Co-like ion and the
$3d_{3/2}$--$3d_{5/2}$ transition in the K-like ion. For the former, our
measured wavelength of 21.229$\pm$0.003~nm agrees within uncertainties with the
semiempirical prediction of 21.202$\pm$0.033~nm \cite{Ekberg_1987}, while the
calculated FAC value is slightly outside the experimental error bars. As also
was the case for the W measurements~\cite{Draganic_etal_Hf_Ta_Au_2011}, the
measured wavelength of 18.158$\pm$0.003~nm for the M1 line in K-like ion
agrees very well with the multiconfiguration Dirac-Fock (MCDF) result of
18.1578~nm obtained by Ali and Kim \cite{Ali_Kim_1992}.

The last column in Table~\ref{TabHf} shows the calculated transition
probabilities for the identified lines of Hf. The M1 lines have A values in
the range of 8$\times$10$^4$ s$^{-1}$ to 3$\times$10$^6$ s$^{-1}$. The 
transition probability for the electric-quadrupole line at 16.149 nm is much
smaller, about 2.8$\times$10$^2$ s$^{-1}$, which exemplifies relative weakness
of intraconfiguration E2 lines as compared to M1 transitions. 

\subsection{Ta spectra}

We indentified 50 new lines from thirteen ions from Ni-like Ta$^{45+}$ to
S-like Ta$^{57+}$ (Table~\ref{TabTa}). The identified lines correspond to
intraconfiguration M1 transitions, although not all originate from the ground
configurations: for instance, all lines in Ar-, Cl-, and S-like ions are
within the $3p^n3d$ configurations with $n$=3--5. Similar to the Hf case, the
MCDF calculations of Ali and Kim agree very well with our wavelength for the
$3d$ line in the K-like ion \cite{Ali_Kim_1992}, and Safronova et al.'s RMBPT
data \cite{Safronova_2007} are very close to the measured wavelength in the
Ni-like ion. The semiempirical value of 19.816$\pm$0.032~nm for the M1 line in the
Co-like ion agrees with our measured wavelength of 19.843$\pm$0.003~nm within the
stated uncertainties.

\subsection{Au spectra}

Forty new spectral lines were identified for ions of Au, from Ni-like
Au$^{51+}$ to K-like Au$^{60+}$ (Table~\ref{TabAu}). The higher ion charges
for the isoelectronic ions of Au, as compared to Ta and Hf, result in smaller
relative wavelength separations between the analogous spectral lines.
Therefore, the lines overlap more often and are more difficult to analyze.
This explains the relatively smaller number of identified lines for Au.
Similar to the Hf and Ta cases, the RMBPT \cite{Safronova_2007} and MCDF
\cite{Ali_Kim_1992} results for Ni-like and K-like ions, respectively, agree
well with the measured wavelengths, and again, the semiempirical wavelength
for the $3d^9$ intraconfiguration transition in the Co-like ion agrees within
uncertainties.

Two electric-quadrupole lines were identified in the measured spectra of Au.
As mentioned above, the transition probabilities for E2 lines are typically
much smaller than for M1 lines, and therefore they are more difficult to
observe due to collisional damping. Indeed, our calculations show that the
A-values for the E2 line at 11.993~nm in the V-like ion and for the E2 line in
the Sc-like ion are only 7.97$\times$10$^2$ s$^{-1}$ and 1.55$\times$10$^3$
s$^{-1}$, respectively, while the A-values for M1 lines are on the order of
$10^6$ s$^{-1}$. The fact that such low-A lines were observed in the EBIT is
due to their high branching ratios: both E2 transitions are the strongest for
their corresponding upper levels.

We also identified two known lines in the Ni-like and Co-like ions of Au. The
newly measured wavelengths of 13.858$\pm$0.003~nm and 13.517$\pm$0.003~nm
agree within uncertainties with our previous results of 13.860$\pm$0.003~nm
and 13.522$\pm$0.003~nm \cite{Draganic_etal_Hf_Ta_Au_2011}. Thus, the
recommended averaged values for these lines are 13.8590$\pm$0.0020~nm for the
$3d^94s$ transition in the Ni-like Au$^{51+}$ and 13.5195$\pm$0.0020~nm for
the $3d^9$ transition in the Co-like Au$^{52+}$.

\section{Conclusions}

In this paper we presented measurements and identifications of a large number
of EUV spectral lines originating from forbidden transitions, mostly within
the ground configurations, in the $3d^n$ ions of Hf, Ta, and Au. These results
provide a new accurate set of atomic data for highly-charged ions of high-Z
elements which are important for magnetic and intertial confinement fusion
research. Moreover, the measured wavelengths can also be used to test the most
advanced theories of atomic structure in the relativistic regime.

As was shown in our previous work~\cite{16374EL}, the intensity ratios for
similar M1 lines in the $3d^n$ ions of W are very sensitive to electron
densities in the parameter range of magnetic fusion plasmas. Since the
physical processes affecting level populations for Hf, Ta and Au are the same
as those for W, one can expect similar sensitivity for the presently measured
lines as well. The only minor difference will be that the sensitivity range of
densities for Hf and Ta ions will be slightly shifted to lower densities while
for Au it will be shifted to higher densities. This results from a strong
$z$-dependence of the M1 radiative rates.

\section{Acknowledgments} 
We are grateful to J. M. Pomeroy, J. N. Tan, and S. M. Brewer for assistance
during the early experimental phase of this work, to U. I. Safronova for
providing the energy levels of Ni-like ions from Ref.~\cite{Safronova_2007},
and to Y.A. Podpaly for valuable comments.
This work is supported in part by the Office of Fusion Energy Sciences of the
U.S. Department of Energy.

%\section{References}

\bibliographystyle{epj}
\bibliography{epj1}{}

\begin{landscape}
\newpage

\begin{figure*}

\includegraphics[width=0.9\textwidth]{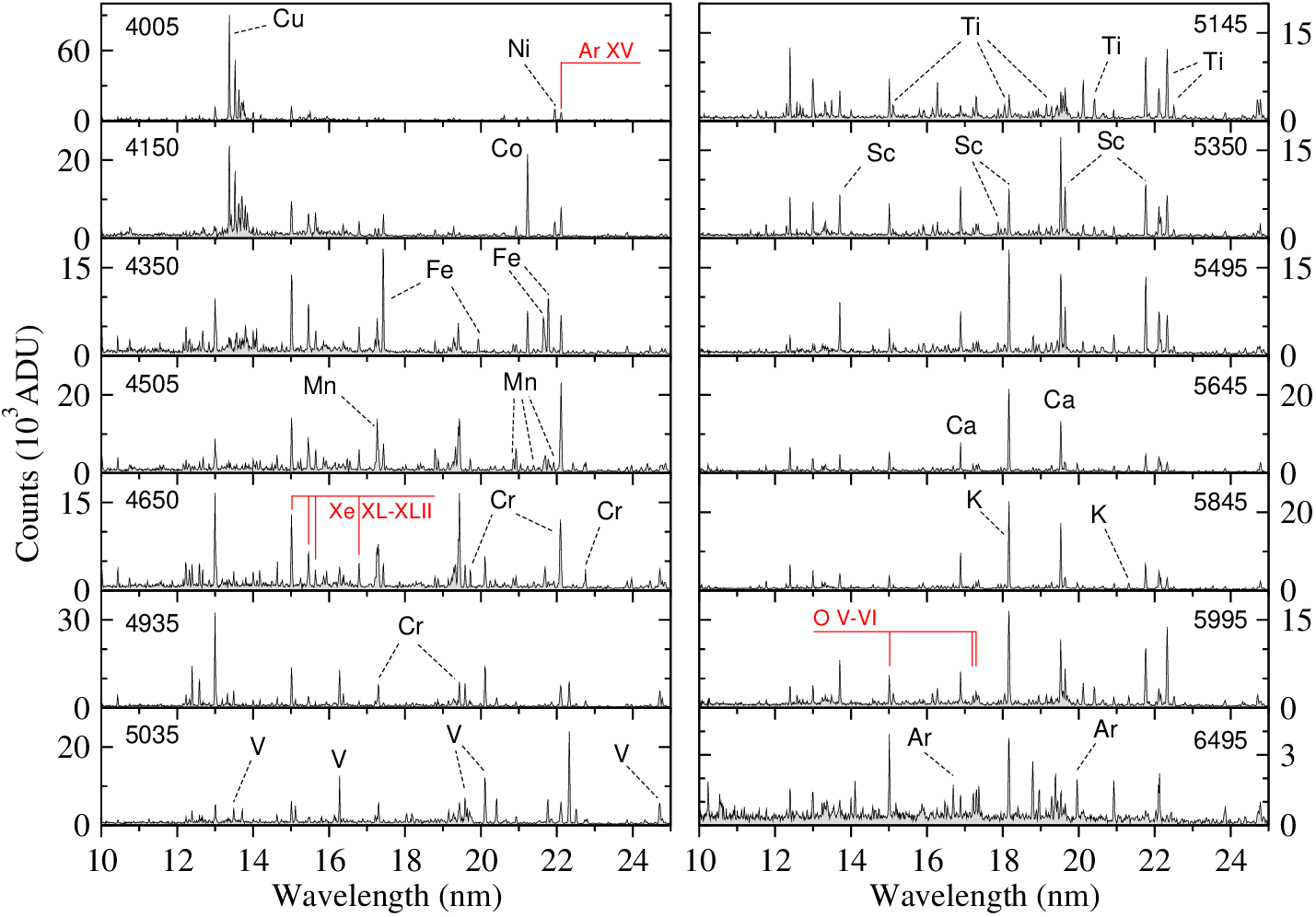}

\caption{ Experimental spectra for Hf. Nominal beam energies (in eV) are
shown in the upper corners. Some strong identified lines are indicated by
their isoelectronic sequences. Several impurity lines from Xe and O are 
indicated as well.}
\label{Fig1}
\end{figure*}

\newpage

\begin{figure*}
\includegraphics[width=0.9\textwidth]{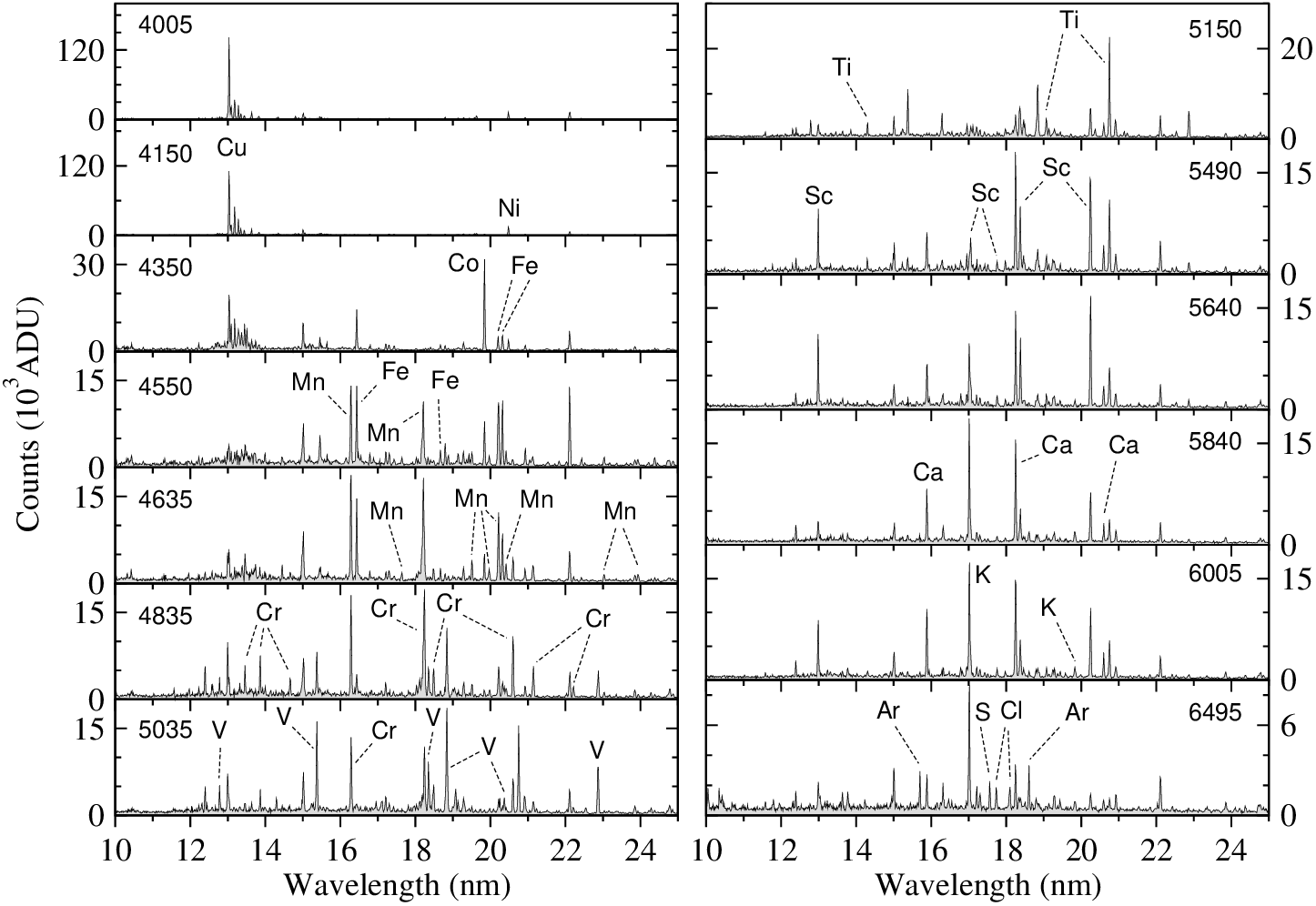}
\caption{ Experimental spectra for Ta. Nominal beam energies (in eV) are
shown in the upper corners. Some strong identified lines are indicated
by their isoelectronic sequences.}

\label{Fig2}
\end{figure*}

\newpage

\begin{figure*}
\includegraphics[width=0.9\textwidth]{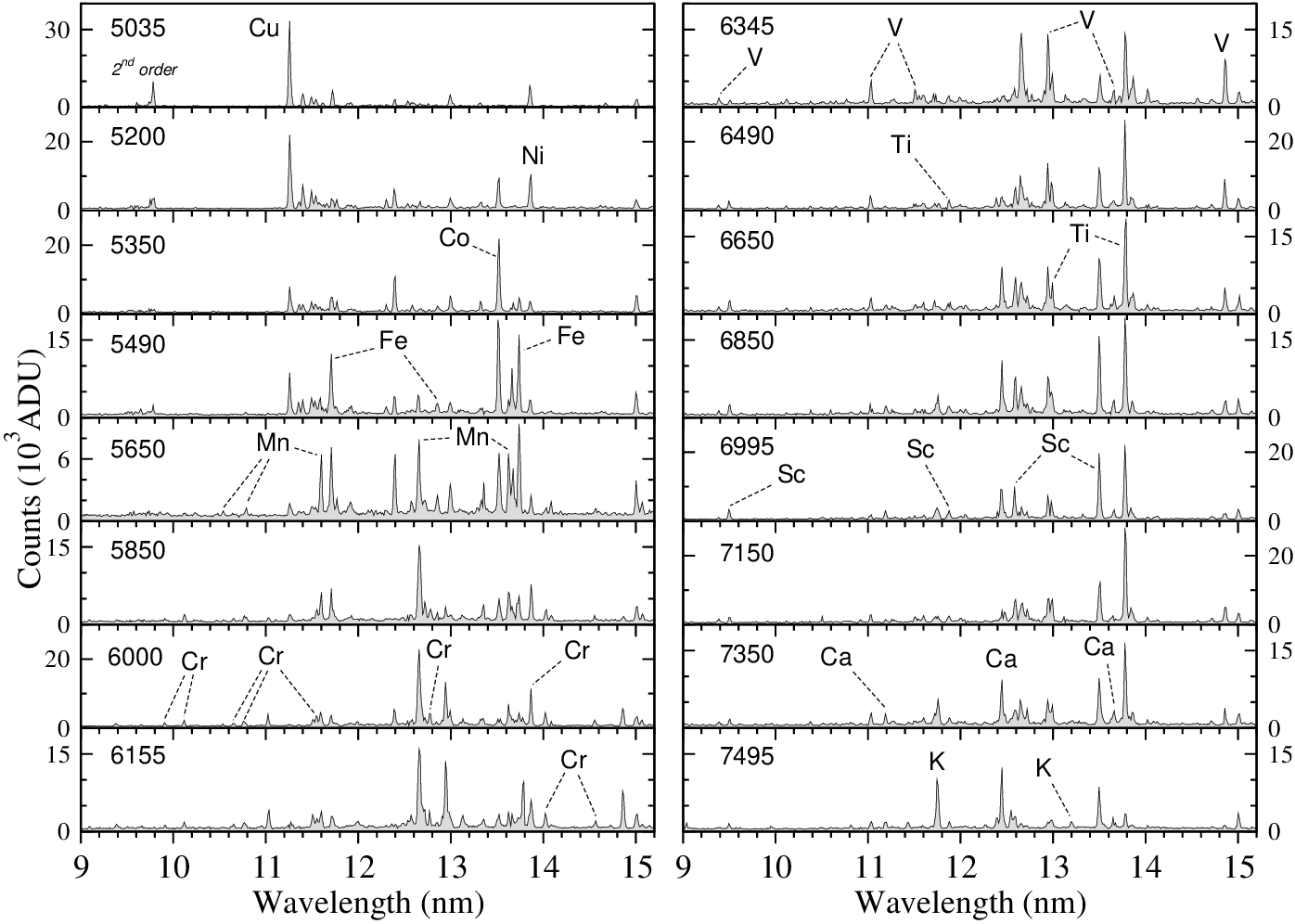}
\caption{ Experimental spectra for Au. Nominal beam energies (in eV) are shown
in the upper corners. Some strong identified lines are indicated by their
isoelectronic sequences.}

\label{Fig3}
\end{figure*}

\end{landscape}

\newpage

\begin{figure*}
\includegraphics[width=0.95\textwidth]{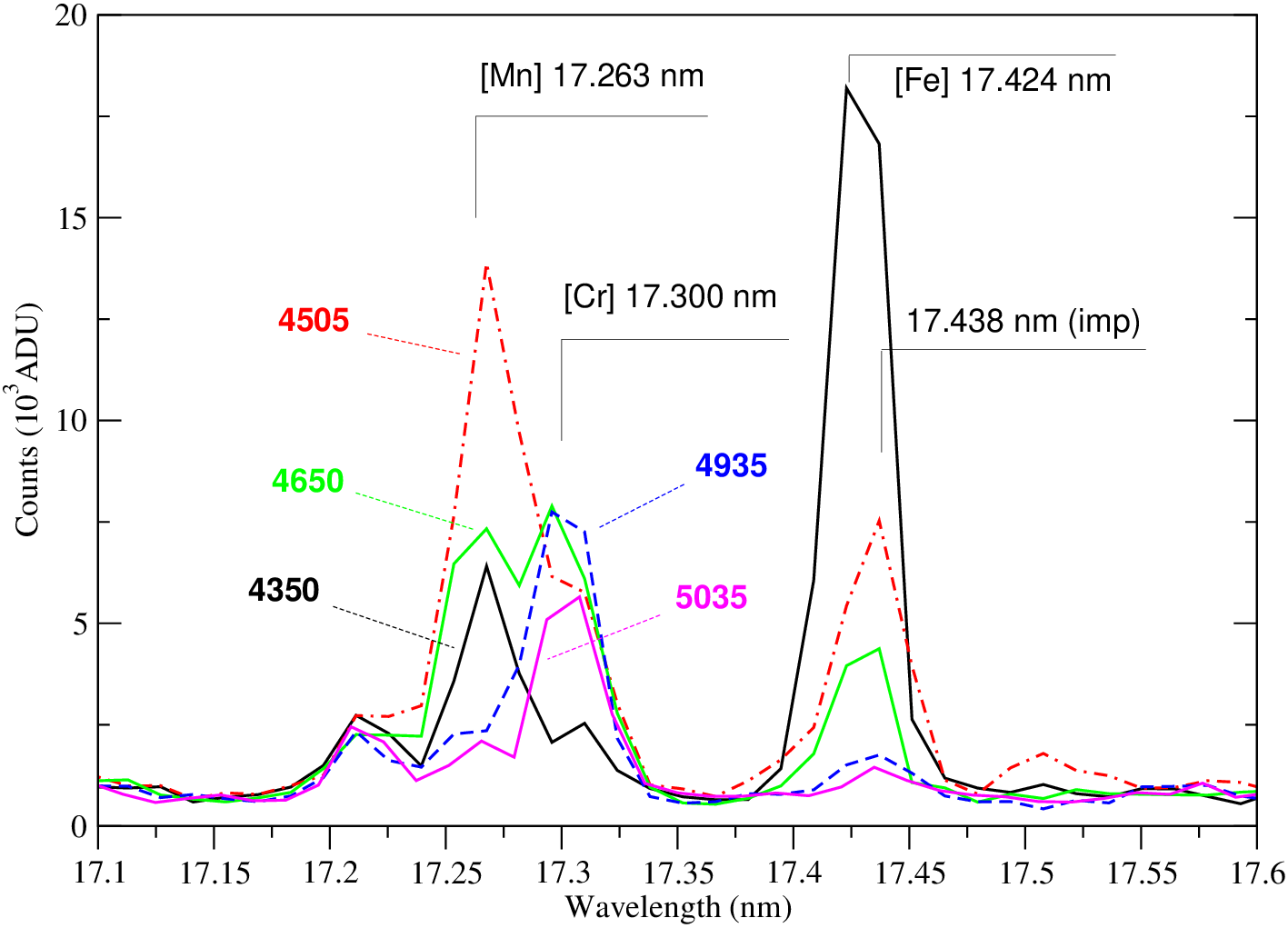}
\caption{Hafnium spectra near 17.3~nm at the nominal beam energies between
4350 eV and 5035 eV. The unknown impurity line at
17.438~nm was prominent in the background spectra.}
\label{Fig1b}
\end{figure*}

\newpage

\begin{table*}
\caption[]{Nominal electron beam energies (in eV) used in the present work.}
\centering
\begin{tabular}{c|cccccccccccccccc}
\hline\noalign{\smallskip}
Hf & 4005 & 4150 &4350 &4505 &4650 &4935 &5035 &5145 &5350 & 5495 &5645 
&5845 &5995 &6495 \\
Ta & 4005 & 4150 & 4350 & 4550 & 4635 &  4835 & 5035 &  5150 & 5490 & 5640 & 5840 & 6005 &
6495 &&&\\
Au & 5035 & 5200 & 5350 & 5490 &5650 & 5850 &6000 & 6155 & 6345 & 6490 & 6650 &
6850 & 6995 & 7150 & 7350 &7495 \\
\noalign{\smallskip}\hline
\end{tabular}
\label{TabEb}
\end{table*}

\newpage

\begin{table*}
\caption[]{Calculated ionization potentials (in eV) of Ni-like through Cl-like ions of Hf, Ta and Au.}
\centering
\begin{tabular}{llcccccc}
\hline\noalign{\smallskip}
\multicolumn{2}{c}{Sequence}  & \multicolumn{2}{c}{Hf} & \multicolumn{2}{c}{Ta} & \multicolumn{2}{c}{Au} \\
     &  & FAC & Ref.~\cite{8672EL} & FAC & Ref.~\cite{8672EL} & FAC & Ref.~\cite{8672EL} \\
\noalign{\smallskip}\hline\noalign{\smallskip}
 Ni & $3d^{10}$ &  3741.7  & 3741 &  3895.3  & 3898.7 &  4880.4  & 4888    \\
 Co & $3d^9$ 	&  3859.4  & 3858 &  4015.2  & 4014   &  5014.0  & 5013    \\
 Fe & $3d^8$ 	&  3984.9  & 3984 &  4143.1  & 4143   &  5156.0  & 5156    \\
 Mn & $3d^7$ 	&  4119.1  & 4118 &  4279.8  & 4278   &  5307.9  & 5307    \\
 Cr & $3d^6$ 	&  4248.2  & 4246 &  4411.3  & 4410   &  5453.7  & 5452    \\
 V  & $3d^5$ 	&  4375.7  & 4372 &  4541.0  & 4537   &  5596.4  & 5594    \\
 Ti & $3d^4$ 	&  4576.9  & 4573 &  4749.1  & 4745   &  5850.8  & 5846    \\
 Sc & $3d^3$ 	&  4709.1  & 4703 &  4883.6  & 4877   &  5999.5  & 5994    \\
 Ca & $3d^2$ 	&  4853.3  & 4846 &  5030.4  & 5024   &  6162.3  & 6156    \\
 K  & $3d  $ 	&  4987.6  & 4980 &  5167.1  & 5159   &  6313.6  & 6305    \\
 Ar & $3p^6$ 	&  5359.8  & 5350 &  5545.9  & 5537   &  6732.4  & 6724    \\
 Cl & $3p^5$ 	&  5474.9  & 5468 &  5663.0  & 5655   &  6861.3  & 6854    \\
\noalign{\smallskip}\hline
\end{tabular}
\label{TabIP}
\end{table*}

\newpage
%\squeezetable
\begin{table*}
\caption[]{Identified spectral lines of highly-charged ions of hafnium.
Level identifications include calculated level numbers. 
Blended lines are indicated by ``\textit{b}". Notation a(b) for transition probabilities $A$ means $a \cdot 10^b$. 
The electric-quadrupole line is marked by asterisk. Other 
theoretical works: A -- \cite{Safronova_2007}, B -- \cite{Ekberg_1987}, C --
\cite{Ali_Kim_1992}.
}
\begin{tabular}{cclllllllll}
\hline\noalign{\smallskip}
 Ion   & Seq. & Conf. & \multicolumn{2}{l}{Lower level} & \multicolumn{2}{l}{Upper
 level} & $\lambda_{exp}$ & $\lambda_{th}$ & A \\
charge &      &  & No. & Term$_J$ &  No. & Term$_J$ &(nm) & (nm) &  (s$^{-1}$) \\
\noalign{\smallskip}\hline\noalign{\smallskip}
44	&	Ni	&	$3d^9 4s$ & 3 & 	 $((d_+^5)_{5/2},s_+)_2$	&	 4 & 	$((d_-^3)_{3/2},s_+)_1$	& 21.944(3)	&	22.0114, 21.9377$^A$ &	1.60(6) \\
45	&	Co	&	$3d^9$ &  1 & 	$(d_+^5)_{5/2}$	&	 2 & 	$(d_-^3)_{3/2}$	& 21.229(3)	&	21.3176, 21.202(33)$^B$		&	1.65(6) \\
46	&	Fe	&	$3d^8$ &  1 & 	$(d_+^4)_4$	&	 7 &   $((d_-^3)_{3/2},(d_+^5)_{5/2})_4$   &   17.424(8)   &   17.3934 &   6.90(5) \\
46	&	Fe	&	$3d^8$ &  2 & 	$(d_+^4)_2$	&	 6 &   $((d_-^3)_{3/2},(d_+^5)_{5/2})_1$   &   19.929(3)   &   19.9516 &   1.15(6)\\
46	&	Fe	&	$3d^8$ &  2 & 	$(d_+^4)_2$	&	 5 &   $((d_-^3)_{3/2},(d_+^5)_{5/2})_2$   &   21.652(5)   &   21.7390 &   1.29(6)\\
46	&	Fe	&	$3d^8$ &  1 & 	$(d_+^4)_4$	&	 4 &   $((d_-^3)_3/2,(d_+^5)_{5/2})_3$	   &   21.775(3)   &   21.8812 &   2.15(6)\\
47	&	Mn	&	$3d^7$ &  1 & 	$(d_+^3)_{9/2}$	&	9 &   $((d_-^3)_{3/2},(d_+^4)_4)_{11/2}$  &   17.263(3)    &  17.2267 &   1.28(5)\\
47	&	Mn	&	$3d^7$ &  1 & 	$(d_+^3)_{9/2}$	&	5 &   $((d_-^3)_{3/2},(d_+^4)_4)_{9/2}$   &   19.403(3)   &   19.4231 &   1.51(6)\\
47	&	Mn	&	$3d^7$ &  3 & 	$(d_+^3)_{5/2}$	&	10 &  $((d_-^3)_{3/2},(d_+^4)_2)_{7/2}$   &   20.851(3)   &   20.8488 &   1.84(5)\\
47	&	Mn	&	$3d^7$ &  2 & 	$(d_+^3)_{3/2}$	&	8 &   $((d_-^3)_{3/2},(d_+^4)_2)_{1/2}$   &   21.390(3)   &   21.4662 &   1.72(6)\\
47	&	Mn	&	$3d^7$ &  1 & 	$(d_+^3)_{9/2}$	&	4 &   $((d_-^3)_{3/2},(d_+^4)_4)_{7/2}$   &   21.687(3)   &   21.7903 &   2.38(6)\\
47	&	Mn	&	$3d^7$ &  2 & 	$(d_+^3)_{3/2}$	&	 7 &  $((d_-^3)_{3/2},(d_+^4)_4)_{5/2}$   &   21.916(3)   &   21.9934 &   6.80(5)\\
48	&	Cr	&	$3d^6$ &  1 & 	$(d_+^2)_4$	&	 10 &  $((d_-^3)_{3/2},(d_+^3)_{3/2})_3$   &   17.300(3)   &   17.2838 &   7.07(5)\\
48	&	Cr	&	$3d^6$ &  1 & 	$(d_+^2)_4$	&	 8 &   $((d_-^3)_{3/2},(d_+^3)_{9/2})_5$   &   19.437(3)   &   19.4248 &   4.35(5)\\
48	&	Cr	&	$3d^6$ &  2 & 	$(d_+^2)_2$	&	 9 &   $((d_-^3)_{3/2},(d_+^3)_{3/2})_2$   &   19.721(3)   &   19.7351 &   1.35(6)\\
48	&	Cr	&	$3d^6$ &  1 & 	$(d_+^2)_4$	&	 5 &   $((d_-^3)_{3/2},(d_+^3)_{9/2})_4$   &   22.091(3)   &   22.1559 &   2.00(6)\\
48	&	Cr	&	$3d^6$ &  1 & 	$(d_+^2)_4$	&	 4 &   $((d_-^3)_{3/2},(d_+^3)_{9/2})_3$   &   22.754(3)   &   22.8747 &   1.69(6)\\
49	&	V	&	$3d^5$ &  1 & 	$(d_+)_{5/2}$	& 	 10 &  $((d_-^3)_{3/2},(d_+^2)_0)_{3/2}$   &   13.487(3)   &   13.4510 &   1.82(5)\\
49	&	V	&	$3d^5$ &  1 & 	$(d_+)_{5/2}$	&	 9 &   $((d_-^3)_{3/2},(d_+^2)_2)_{7/2}$   &   16.275(3)   &   16.2253 &   8.31(4)\\
49	&	V	&	$3d^5$ &  1 & 	$(d_+)_{5/2}$	&	 5 &   $((d_-^3)_{3/2},(d_+^2)_2)_{3/2}$   &   19.583(3)   &   19.6141 &   2.54(6)\\
49	&	V	&	$3d^5$ &  1 & 	$(d_+)_{5/2}$	&	 3 &   $((d_-^3)_{3/2},(d_+^2)_4)_{7/2}$   &   20.114(3)\textit{b}  &   20.1278 &   1.05(6)\\
%49	&	V	&	$3d^5$ &  4 & 	$((d_-^3)_{3/2},(d_+^2)_4)_{11/2}$	  & 15 & 	$((d_-^2)_2,(d_+^3)_{9/2})_{13/2}$	&	20.114(3)b &	20.1289	& 3.44(5)\\
49	&	V	&	$3d^5$ &  1 & 	$(d_+)_{5/2}$	& 2 & 	$((d_-^3)_{3/2},(d_+^2)_4)_{5/2}$	&	24.711(3)	&	24.8671	&	2.19(6)\\
50	&	Ti	&   $3d^4$ &  5 & 	$((d_-^3)_{3/2},d_+)_3$	&	 17 &  $((d_-^2)_0,(d_+^2)_4)_4$   &   15.105(3)   &   15.0733 &   7.48(5)\\
50	&	Ti	&	$3d^4$ &  3 & 	$((d_-^3)_{3/2},d_+)_4$	&    13 &  $((d_-^2)_2,(d_+^2)_2)_3$   &   18.043(3)   &   18.0451 &   1.19(6)\\
50	&	Ti	&	$3d^4$ &  2 & 	$((d_-^3)_{3/2},d_+)_1$	&	 7 &   $((d_-^2)_2,(d_+^2)_4)_2$   &   19.148(3)   &   19.1681 &   3.13(6)\\
50	&	Ti	&	$3d^4$ &  3 & 	$((d_-^3)_{3/2},d_+)_4$	&	 10 &  $((d_-^2)_2,(d_+^2)_4)_5$   &   20.410(3)   &   20.4503 &   1.10(6)\\
50	&	Ti	&	$3d^4$ &  1 & 	$(d_-^4)_0$	&	             2 &   $((d_-^3)_{3/2},d_+)_1$ &   22.325(3)   &   22.4327 &   2.17(6)\\
50	&	Ti	&	$3d^4$ &  3 & 	$((d_-^3)_{3/2},d_+)_4$	&	 8 &   $((d_-^2)_2,(d_+^2)_4)_4$   &   22.502(3)   &   22.6346 &   1.98(6)\\
51	&	Sc	&	$3d^3$ &  1 & 	$(d_-^3)_{3/2}$	&	         7 &   $((d_-^2)_0,d_+)_{5/2}$ &   13.706(3)   &   13.6749 &   1.92(5)\\
51	&	Sc	&	$3d^3$ &  5 & 	$((d_-^2)_2,d_+)_{9/2}$	&	 12 &  $(d_-,(d_+^2)_4)_{11/2}$    &   17.873(3)   &   17.8904 &   9.60(5)\\
51	&	Sc	&	$3d^3$ &  1 & 	$(d_-^3)_{3/2}$	&	         6 &   $((d_-^2)_2,d_+)_{1/2}$ &   18.150(6)   &   18.1838 &   6.84(5)\\
51	&	Sc	&	$3d^3$ &  2 & 	$((d_-^2)_2,d_+)_{5/2}$	&	 8 &   $(d_-,(d_+^2)_4)_{7/2}$ &   18.942(3)   &   18.9797 &   3.29(6)\\
51	&	Sc	&	$3d^3$ &  1 & 	$(d_-^3)_{3/2}$	&	         3 &   $((d_-^2)_2,d_+)_{3/2}$ &   19.639(3)   &   19.6876 &   1.83(6)\\
51	&	Sc	&	$3d^3$ &  1 & 	$(d_-^3)_{3/2}$	&	         2 &   $((d_-^2)_2,d_+)_{5/2}$ &   21.762(3)	&  21.8713 &   2.26(6)\\
52	&	Ca	&	$3d^2$ &  1 & 	$(d_-^2)_2$	&	 5 &   $(d_-,d_+)_4$   &   16.149(4)*   &   16.1575 &   2.82(2)\\
52	&	Ca	&	$3d^2$ &  1 & 	$(d_-^2)_2$	&	 4 &   $(d_-,d_+)_2$   &   16.884(3)   &   16.9070 &   1.23(6)\\
52	&	Ca	&	$3d^2$ &  1 & 	$(d_-^2)_2$	&	 3 &   $(d_-,d_+)_3$   &   19.525(3)   &   19.6012 &   2.48(6)\\
52	&	Ca	&	$3d^2$ &  2 & 	$(d_-^2)_0$	&	 6 & 	$(d_-,d_+)_1$	&	22.158(3)	&	22.2673 &	1.13(6)\\
53	&	K	&	$3p^5 3d^2$	& 6 & 	$((p_+^3)_{3/2},(d_-^2)_2)_{7/2}$	&	 19 & 	$((p_+^3)_{3/2},d_-d_+)_{9/2}$	&	14.576(3)	&	14.5240	&	8.75(4)\\
53	&	K	&	$3d$	& 1 & 	$(d_-)_{3/2}$	&	 2 & 	$(d_+)_{5/2}$	&	18.158(3)	&	18.2132, 18.1578$^{C}$	&	1.76(6)\\
53	&	K	&	$3p^5 3d^2$	& 6 & 	$((p_+^3)_{3/2},(d_-^2)_2)_{7/2}$	&	 9 & 	$((p_+^3)_{3/2}d_-d_+)_{9/2}$	&	21.316(3)	&	21.4145	&	1.96(6)\\
54	&	Ar	&	$3p^5 3d$	& 5 & 	$((p_+^3)_{3/2},d_-)_2$	&	 8 & 	$((p_+^3)_{3/2},d_+)_3$	&	16.693(3)	&	16.7008	&	1.76(6)\\
54	&	Ar	&	$3p^5 3d$	& 4 & 	$((p_+^3)_{3/2},d_-)_3$	&	 6 & 	$((p_+^3)_{3/2},d_+)_4$	&	19.958(3)	&	20.0564	&	1.39(6)\\

\noalign{\smallskip}\hline
%\end{tabular}
\end{tabular}
\label{TabHf}
\end{table*}

\newpage
%\squeezetable
%\begin{landscape}
\begin{table*}
\caption[]{Identified spectral lines of highly-charged ions of tantalum.
Level identifications include calculated level numbers. 
Blended lines are indicated by ``\textit{b}". Notation a(b) for transition probabilities $A$ means $a \cdot 10^b$. 
Other 
theoretical works: A -- \cite{Safronova_2007}, B -- \cite{Ekberg_1987}, C --
\cite{Ali_Kim_1992}.
}
\begin{tabular}{cclllllllll}
\hline\noalign{\smallskip}
 Ion   & Seq. & Conf. & \multicolumn{2}{l}{Lower level} & \multicolumn{2}{l}{Upper
 level} & $\lambda_{exp}$ & $\lambda_{th}$ & A \\
charge &      &  & No. & Term$_J$ &  No. & Term$_J$ &(nm) & (nm) &  (s$^{-1}$) \\
\noalign{\smallskip}\hline\noalign{\smallskip}
\label{TabTa}
45	&	Ni	&	$3d^9 4s$ & 3 &	$((d_+^5)_{5/2},s_+)_2$	&	 4 &	$((d_-^3)_{3/2},s_+)_1$	&	20.482(3)	&	20.5428, 20.4758$^A$	&	1.96(6) \\
46	&	Co	&	$3d^9$ & 1 &	$(d_+^5)_{5/2}$	&	 2 &	$(d_-^3)_{3/2}$	&	19.843(3)	&	19.9229, 19.816(32)$^B$	&	2.02(6)\\
47	&	Fe	&	$3d^8$ & 1 &	$(d_+^4)_4$	&	 7 &	$((d_-^3)_{3/2},(d_+^5)_{5/2})_4$	&	16.436(3)	&	16.4099	&	8.37(5)\\
47	&	Fe	&	$3d^8$ & 2 &	$(d_+^4)_2$	&	 6 &	$((d_-^3)_{3/2},(d_+^5)_{5/2})_1$	&	18.667(3)	&	18.6907	&	1.40(6)\\
47	&	Fe	&	$3d^8$ & 2 &	$(d_+^4)_2$	&	5 &	$((d_-^3)_{3/2},(d_+^5)_{5/2})_2$	&	20.204(3)	&	20.2858	&	1.59(6)\\
47	&	Fe	&	$3d^8$ & 1 &	$(d_+^4)_4$	&	4 &	$((d_-^3)_{3/2},(d_+^5)_{5/2})_3$	&	20.322(3)	&	20.4182	&	2.64(6)\\
48	&	Mn	&	$3d^7$ & 1 &	$(d_+^3)_{9/2}$	&	 9 &	$((d_-^3)_{3/2},(d_+^4)_4)_{11/2}$	&	16.281(4)	&	16.2525	&	1.56(5)\\
48	&	Mn	&	$3d^7$ & 5 &	$((d_-^3)_{3/2},(d_+^4)_4)_{9/2}$	&	 15 &	$((d_-^2)_2,(d_+^5)_{5/2})_{9/2}$	&	17.640(3)	&	17.6608	&	1.81(6)\\
48	&	Mn	&	$3d^7$ & 3 &	$(d_+^3)_{5/2}$	&	 11 &	$((d_-^3)_{3/2},(d_+^4)_2)_{5/2}$	&	18.169(3)	&	18.1220	&	7.02(5)\\
48	&	Mn	&	$3d^7$ & 1 &	$(d_+^3)_{9/2}$	&	 5 &	$((d_-^3)_{3/2},(d_+^4)_4)_{9/2}$	&	18.210(3)	&	18.2328	&	1.85(6)\\
48	&	Mn	&	$3d^7$ & 3 &	$(d_+^3)_{5/2}$	&	 10 &	$((d_-^3)_{3/2},(d_+^4)_2)_{7/2}$	&	19.508(3)	&	19.5120	&	2.28(5)\\
48	&	Mn	&	$3d^7$ & 2 &	$(d_+^3)_{3/2}$	&	 8 &	$((d_-^3)_{3/2},(d_+^4)_2)_{1/2}$	&	19.968(3)	&	20.0415	&	2.11(6)\\
48	&	Mn	&	$3d^7$ & 1 &	$(d_+^3)_{9/2}$	&	 4 &	$((d_-^3)_{3/2},(d_+^4)_4)_{7/2}$	&	20.215(3)	&	20.3167	&	2.92(6)\\
48	&	Mn	&	$3d^7$ & 2 &	$(d_+^3)_{3/2}$	&	 7 &	$((d_-^3)_{3/2},(d_+^4)_4)_{5/2}$	&	20.417(3)	&	20.4889	&	8.38(5)\\
48	&	Mn	&	$3d^7$ & 3 &	$(d_+^3)_{5/2}$	&	 7 &	$((d_-^3)_{3/2},(d_+^4)_4)_{5/2}$	&	23.031(3)	&	23.1203	&	7.00(5)\\
48	&	Mn	&	$3d^7$ & 3 &	$(d_+^3)_{5/2}$	&	 6 &	$((d_-^3)_{3/2},(d_+^4)_2)_{3/2}$	&	23.938(3)\textit{b}	&	24.0887	&	1.38(6)\\
49	&	Cr	&	$3d^6$ & 1 &	$(d_+^2)_4$	&	 15 &	$((d_-^3)_{3/2},(d_+^3)_{5/2})_3$	&	13.462(3)	&	13.3992	&	3.39(5)\\
49	&	Cr	&	$3d^6$ & 1 &	$(d_+^2)_4$	&	 12 &	$((d_-^3)_{3/2},(d_+^3)_{5/2})_4$	&	13.861(3)	&	13.8207	&	3.13(4)\\
49	&	Cr	&	$3d^6$ & 2 &	$(d_+^2)_2$	&	 15 &	$((d_-^3)_{3/2},(d_+^3)_{5/2})_3$	&	14.665(3)	&	14.6071	&	4.07(5)\\
49	&	Cr	&	$3d^6$ & 1 &	$(d_+^2)_4$ &	 10 &	$((d_-^3)_{3/2},(d_+^3)_{3/2})_3$	&	16.286(4)	&	16.2870	&	8.47(5)\\
49	&	Cr	&	$3d^6$ & 1 &	$(d_+^2)_4$	&	 8 &	$((d_-^3)_{3/2},(d_+^3)_{9/2})_5$	&	18.242(5)	&	18.2370	&	5.35.(5)\\
49	&	Cr	&	$3d^6$ & 2 &	$(d_+^2)_2$	&	 9 &	$((d_-^3)_{3/2},(d_+^3)_{3/2})_2$	&	18.486(3)	&	18.5076	&	1.66(6)\\
49	&	Cr	&	$3d^6$ & 1 &	$(d_+^2)_4$	&	 5 &	$((d_-^3)_{3/2},(d_+^3)_{9/2})_4$	&	20.602(3)	&	20.6685	&	2.46(6)\\
49	&	Cr	&	$3d^6$ & 1 &	$(d_+^2)_4$	&	 4 &	$((d_-^3)_{3/2},(d_+^3)_{9/2})_3$	&	21.144(3)	&	21.2545	&	2.08(6)\\
49	&	Cr	&	$3d^6$ & 2 &	$(d_+^2)_2$	&	 6 &	$((d_-^3)_{3/2},(d_+^3)_{3/2})_1$	&	22.214(3)	&	22.3583	&	2.66(6)\\
50	&	V	&	$3d^5$ & 1 &	$(d_+)_{5/2}$	&	 10 &	$((d_-^3)_{3/2},(d_+^2)_0)_{3/2}$	&	12.787(3)	&	12.7531	&	2.16(5)\\
50	&	V	&	$3d^5$ & 1 &	$(d_+)_{5/2}$	&	 9 &	$((d_-^3)_{3/2},(d_+^2)_2)_{7/2}$	&	15.374(3)	&	15.3315	&	1.00(5)\\
50	&	V	&	$3d^5$ & 1 &	$(d_+)_{5/2}$	&	 5 &	$((d_-^3)_{3/2},(d_+^2)_2)_{3/2}$	&	18.352(3)	&	18.3840	&	3.09(6)\\
50	&	V	&	$3d^5$ & 1 &	$(d_+)_{5/2}$	&	 3 &	$((d_-^3)_{3/2},(d_+^2)_4)_{7/2}$	&	18.841(3)\textit{b}	&	18.8579	&	1.30(6)\\
50	&	V	&	$3d^5$ & 4 &	$((d_-^3)_{3/2},(d_+^2)_4)_{11/2}$	&	 15 &	$((d_-^2)_2,(d_+^3)_{9/2})_{13/2}$	&	18.841(3)\textit{b} 	&	18.8432	&	4.23(5)\\
50	&	V	&	$3d^5$ & 4 &	$((d_-^3)_{3/2},(d_+^2)_4)_{11/2}$	&	 13 &	$((d_-^2)_2,(d_+^3)_{9/2})_{11/2}$	&	20.368(3)	&	20.4465	&	1.88(6)\\
50	&	V	&	$3d^5$ & 1 &	$(d_+)_{5/2}$	&	 2 &	$((d_-^3)_{3/2},(d_+^2)_4)_{5/2}$	&	22.869(3)	&	23.0137	&	2.73(6)\\
51	&	Ti	&	$3d^4$ & 5 &	$((d_-^3)_{3/2},d_+)_3$	&	 17 &	$((d_-^2)_0,(d_+^2)_4)_4$	&	14.303(3)	&	14.2704	&	9.03(5)\\
51	&	Ti	&	$3d^4$ & 3 &	$((d_-^3)_{3/2},d_+)_4$	&	 10 &	$((d_-^2)_2,(d_+^2)_4)_5$	&	19.073(3)	&	19.1147	&	1.35(6)\\
51	&	Ti	&	$3d^4$ & 1 &	$(d_-^4)_0$	&	 2 &	$((d_-^3)_{3/2},d_+)_1$	&	20.750(3)	&	20.8492	&	2.69(6)\\
52	&	Sc	&	$3d^3$ & 1 &	$(d_-^3)_{3/2}$	&	 7 &	$((d_-^2)_0,d_+)_{5/2}$	&	12.987(7)	&	12.9603	&	2.31(5)\\
52	&	Sc	&	$3d^3$ & 1 &	$(d_-^3)_{3/2}$	&	 6 &	$((d_-^2)_2,d_+)_{1/2}$	&	17.053(3)	&	17.0818	&	8.37(5)\\
52	&	Sc	&	$3d^3$ & 2 &	$((d_-^2)_2,d_+)_{5/2}$	&	 8 &	$(d_-,(d_+^2)_4)_{7/2}$	&	17.758(3)	&	17.8065	&	4.01(6)\\
52	&	Sc	&	$3d^3$ & 1 &	$(d_-^3)_{3/2}$	&	 3 &	$((d_-^2)_2,d_+)_{3/2}$	&	18.377(3)	&	18.4346	&	2.24(6)\\
52	&	Sc	&	$3d^3$ & 1 &	$(d_-^3)_{3/2}$	&	 2 &	$((d_-^2)_2,d_+)_{5/2}$	&	20.247(3)	&	20.3630	&	2.78(6)\\
53	&	Ca	&	$3d^2$ & 1 &	$(d_-^2)_2$	&	 4 &	$(d_-,d_+)_2$	&	15.885(3)	&	15.9086	&	1.49(6)\\
53	&	Ca	&	$3d^2$ & 1 &	$(d_-^2)_2$	&	 3 &	$(d_-,d_+)_3$	&	18.250(3)	&	18.3192	&	3.02(6)\\
53	&	Ca	&	$3d^2$ & 2 &	$(d_-^2)_0$	&	 6 &	$(d_-,d_+)_1$	&	20.598(3)	&	20.6941	&	1.40(6)\\
54	&	K	&	$3d$ & 1 &	$(d_-)_{3/2}$	&	 2 &	$(d_+)_{5/2}$	&	17.015(3)	&	17.0662, 17.0145$^C$	&	2.14(6)\\
54	&	K	&	$3p^5 3d^2$ & 6 &	$((p_+^3)_{3/2},(d_-^2)_2)_{7/2}$	&	 9 &	$((p_+^3)_{3/2},d_-,d_+)_{9/2}$	&	19.831(3)	&	19.9134	&	2.43(6)\\
55	&	Ar	&	$3p^5 3d$ & 5 &	$((p_+^3)_{3/2},d_-)_2$	&	 8 & $((p_+^3)_{3/2},d_+)_3$ &   15.699(3)   &   15.7090 &   2.12(6)\\
55	&	Ar	&	$3p^5 3d$ & 4 &	$((p_+^3)_{3/2},d_-)_3$	&	 6 & $((p_+^3)_{3/2},d_+)_4$ &   18.607(3)   &   18.6975 &   1.71(6)\\
56	&	Cl	&	$3p^4 3d$	& 3 &	$((p_+^2)_2,d_-)_{5/2}$	&	 7 &   $((p_+^2)_2,d_+)_{7/2}$ &   17.735(3)   &   17.7940 &   1.51(6)\\
56	&	Cl	&	$3p^4 3d$	& 5 &	$((p_+^2)_2,d_-)_{7/2}$	&	 8 &   $((p_+^2)_2,d_+)_{9/2}$ &   18.093(3)   &   18.1197 &   1.88(6)\\
57	&	S	&	$3p^3 3d$ & 6 &	$(p_+,d_-)_3$	&	 7 &	$(p_+,d_+)_4$	&	17.561(3)	&	17.5900	&	2.14(6)\\
\noalign{\smallskip}\hline
\end{tabular}
\end{table*}
%\end{landscape}

\newpage
%\squeezetable
\begin{table*}
\caption[]{Identified spectral lines of highly-charged ions of gold.
Level identifications include calculated level numbers. 
Blended lines are indicated by ``\textit{b}". Notation a(b) for transition probabilities $A$ means $a \cdot 10^b$. 
Electric-quadrupole lines are marked by asterisks. Other 
theoretical works: A -- \cite{Safronova_2007}, B -- \cite{Ekberg_1987}, C --
\cite{Ali_Kim_1992}.
}
\begin{tabular}{cclllllllll}
\hline\noalign{\smallskip}
 Ion   & Seq. & Conf. & \multicolumn{2}{l}{Lower level} & \multicolumn{2}{l}{Upper
 level} & $\lambda_{exp}$ & $\lambda_{th}$ & A \\
charge &      &  & No. & Term$_J$ &  No. & Term$_J$ &(nm) & (nm) &  (s$^{-1}$) \\
\noalign{\smallskip}\hline\noalign{\smallskip}
\label{TabAu}
51	&	Ni	&	$3d^9 4s$ & 3 &	$((d_+^5)_{5/2},s_+)_2$	&	 4 &	$((d_-^3)_{3/2},s_+)_1$	&	13.858(3)	&	13.8971, 13.8550$^A$	&	6.23(6) \\
52	&	Co	&	$3d^9$ & 1 &	$(d_+^5)_{5/2}$	&	 2 &	$(d_-^3)_{3/2}$	&	13.517(3)	&	13.5678, 13.497(27)$^B$	&	6.38(6)\\
53	&	Fe	&	$3d^8$ & 1 &	$(d_+^4)_4$	&	 6 &	$((d_-^3)_{3/2},(d_+^5)_{5/2})_4$	&	11.707(3)	&	11.6996	&	2.52(6)\\
53	&	Fe	&	$3d^8$ & 2 &	$(d_+^4)_2$	&	 7 &	$((d_-^3)_{3/2},(d_+^5)_{5/2})_1$	&	12.854(3)	&	12.8748	&	4.38(6)\\
53	&	Fe	&	$3d^8$ & 1 &	$(d_+^4)_4$	&	 4 &	$((d_-^3)_{3/2},(d_+^5)_{5/2})_3$	&	13.739(3)	&	13.7943	&	8.38(6)\\
53	&	Fe	&	$3d^8$ & 3 &	$(d_+^4)_0$	&	 7 &	$((d_-^3)_{3/2},(d_+^5)_{5/2})_1$	&	16.610(3)	&	16.6990	&	1.86(6)\\
54	&	Mn	&	$3d^7$ & 3 &	$(d_+^3)_{5/2}$	&	 12 &	$((d_-^3)_{3/2},(d_+^4)_0)_{3/2}$	&	10.541(5)	&	10.5155	&	2.49(6)\\
54	&	Mn	&	$3d^7$ & 1 &	$(d_+^3)_{9/2}$	&	 10 &	$((d_-^3)_{3/2},(d_+^4)_2)_{7/2}$	&	10.783(3)	&	10.7761	&	3.89(5)\\
54	&	Mn	&	$3d^7$ & 1 &	$(d_+^3)_{9/2}$	&	 9 &	$((d_-^3)_{3/2},(d_+^4)_4)_{11/2}$	&	11.598(3)	&	11.5917	&	4.80(5)\\
54	&	Mn	&	$3d^7$ & 1 &	$(d_+^3)_{9/2}$	&	 5 &	$((d_-^3)_{3/2},(d_+^4)_4)_{9/2}$	&	12.650(3)\textit{b}	&	12.6728	&	5.70(6)\\
54	&	Mn	&	$3d^7$ & 1 &	$(d_+^3)_{9/2}$	&	 4 &	$((d_-^3)_{3/2},(d_+^4)_4)_{7/2}$	&	13.627(3)	&	13.6811	&	9.29(6)\\
55	&	Cr	&	$3d^6$ & 1 &	$(d_+^2)_4$	&	 15 &	$((d_-^3)_{3/2},(d_+^3)_{5/2})_3$	&	9.910(3)	&	9.8848	&	9.56(5)\\
55	&	Cr	&	$3d^6$ & 1 &	$(d_+^2)_4$	&	 12 &	$((d_-^3)_{3/2},(d_+^3)_{5/2})_4$	&	10.117(3)	&	10.0994	&	8.98(5)\\
55	&	Cr	&	$3d^6$ & 2 &	$(d_+^2)_2$	&	 15 &	$((d_-^3)_{3/2},(d_+^3)_{5/2})_3$	&	10.652(3)	&	10.6295	&	1.22(6)\\
55	&	Cr	&	$3d^6$ & 2 &	$(d_+^2)_2$	&	 14 &	$((d_-^3)_{3/2},(d_+^3)_{5/2})_2$	&	10.764(3)	&	10.755	&	7.40(5)\\
55	&	Cr	&	$3d^6$ & 1 &	$(d_+^2)_4$	&	 10 &	$((d_-^3)_{3/2},(d_+^3)_{3/2})_3$	&	11.553(3)	&	11.5501	&	2.39(6)\\
55	&	Cr	&	$3d^6$ & 1 &	$(d_+^2)_4$	&	 8 &	$((d_-^3)_{3/2},(d_+^3)_{9/2})_5$	&	12.667(3)	&	12.6774	&	1.73(6)\\
55	&	Cr	&	$3d^6$ & 2 &	$(d_+^2)_2$	&	 9 &	$((d_-^3)_{3/2},(d_+^3)_{3/2})_2$	&	12.773(3)	&	12.7997	&	5.20(6)\\
55	&	Cr	&	$3d^6$ & 1 &	$(d_+^2)_4$	&	 5 &	$((d_-^3)_{3/2},(d_+^3)_{9/2})_4$	&	13.869(3)	&	13.9155	&	7.98(6)\\
55	&	Cr	&	$3d^6$ & 1 &	$(d_+^2)_4$	&	 4 &	$((d_-^3)_{3/2},(d_+^3)_{9/2})_3$	&	14.025(3)	&	14.0868	&	6.85(6)\\
55	&	Cr	&	$3d^6$ & 2 &	$(d_+^2)_2$	&	 6 &	$((d_-^3)_{3/2},(d_+^3)_{3/2})_1$	&	14.562(3)	&	14.6410	&	8.98(6)\\
56	&	V	&	$3d^5$ & 1 &	$(d_+)_{5/2}$	&	 10 &	$((d_-^3)_{3/2},(d_+^2)_0)_{3/2}$	&	9.387(3)	&	9.3698	&	5.75(5)\\
56	&	V	&	$3d^5$ & 1 &	$(d_+)_{5/2}$	&	 9 &	$((d_-^3)_{3/2},(d_+^2)_2)_{7/2}$	&	11.028(3)	&	11.0158	&	2.96(5)\\
56	&	V	&	$3d^5$ & 1 &	$(d_+)_{5/2}$	&	 7 &	$((d_-^3)_{3/2},(d_+^2)_2)_{5/2}$	&	11.509(3)	&	11.5065	&	9.17(4)\\
56	&	V	&	$3d^5$ & 1 &	$(d_+)_{5/2}$	&	 6 &	$((d_-^3)_{3/2},(d_+^2)_4)_{9/2}$	&	11.993(3)*	&	11.9941	&	7.97(2)\\
56	&	V	&	$3d^5$ & 1 &	$(d_+)_{5/2}$	&	 5 &	$((d_-^3)_{3/2},(d_+^2)_2)_{3/2}$	&	12.647(3)\textit{b}	&	12.6749	&	9.52(6)\\
56	&	V	&	$3d^5$ & 1 &	$(d_+)_{5/2}$	&	 3 &	$((d_-^3)_{3/2},(d_+^2)_4)_{7/2}$	&	12.947(3)	&	12.9689	&	4.22(6)\\
56	&	V	&	$3d^5$ & 4 &	$((d_-^3)_{3/2},(d_+^2)_4)_{11/2}$	&	 13 &	$((d_-^2)_2,(d_+^3)_{9/2})_{11/2}$	&	13.652(5)	&	13.7051	&	6.11(6)\\
56	&	V	&	$3d^5$ & 1 &	$(d_+)_{5/2}$	&	 2 &	$((d_-^3)_{3/2},(d_+^2)_4)_{5/2}$	&	14.862(3)	&	14.9435	&	9.34(6)\\
57	&	Ti	&	$3d^4$ & 3 &	$((d_-^3)_{3/2},d_+)_4$	&	 13 &	$((d_-^2)_2,(d_+^2)_2)_3$	&	11.870(4)	&	11.8770	&	4.39(6)\\
57	&	Ti	&	$3d^4$ & 3 &	$((d_-^3)_{3/2},d_+)_4$	&	 10 &	$((d_-^2)_2,(d_+^2)_4)_5$	&	12.991(3)	&	13.0196	&	4.30(6)\\
57	&	Ti	&	$3d^4$ & 1 &	$(d_-^4)_0$	&	 2 &	$((d_-^3)_{3/2},d_+)_1$	&	13.781(3)	&	13.8384	&	8.89(6)\\
58	&	Sc	&	$3d^3$ & 1 &	$(d_-^3)_{3/2}$	&	 7 &	$((d_-^2)_0,d_+)_{5/2}$	&	9.502(4)	&	9.4918	&	6.44(5)\\
58	&	Sc	&	$3d^3$ & 1 &	$(d_-^3)_{3/2}$	&	 6 &	$((d_-^2)_2,d_+)_{1/2}$	&	11.887(5)	&	11.9000	&	2.65(6)\\
58	&	Sc	&	$3d^3$ & 1 &	$(d_-^3)_{3/2}$	&	 5 &	$((d_-^2)_2,d_+)_{7/2}$	&	12.054(3)*	&	12.0761	&	1.55(3)\\
58	&	Sc	&	$3d^3$ & 1 &	$(d_-^3)_{3/2}$	&	 3 &	$((d_-^2)_2,d_+)_{3/2}$	&	12.590(3)	&	12.6212	&	7.14(6)\\
58	&	Sc	&	$3d^3$ & 1 &	$(d_-^3)_{3/2}$	&	 2 &	$((d_-^2)_2,d_+)_{5/2}$	&	13.501(3)	&	13.5597	&	8.98(6)\\
59	&	Ca	&	$3d^2$ & 1 &	$(d_-^2)_2$	&	 4 &	$(d_-,d_+)_2$	&	11.189(3)	&	11.2089	&	4.49(6)\\
59	&	Ca	&	$3d^2$ & 1 &	$(d_-^2)_2$	&	 3 &	$(d_-,d_+)_3$	&	12.444(3)	&	12.4874	&	9.35(6)\\
59	&	Ca	&	$3d^2$ & 2 &	$(d_-^2)_0$	&	 6 &	$(d_-,d_+)_1$	&	13.656(3)	&	13.7184	&	4.65(6)\\
60	&	K	&	$3d$ & 1 &	$(d_-)_{3/2}$	&	 2 &	$(d_+)_{5/2}$	&	11.750(8)\textit{b}	&	11.7861, 11.7510$^C$	&	6.49(6)\\
60	&	K	&	$3p^5 3d^2$ & 6 &	$((p_+^3)_{3/2},(d_-^2)_2)_{7/2}$	&	 9 &	$((p_+^3)_{3/2},d_-)_{9/2}$	&	13.200(3)	&	13.2556	&	7.95(6)\\
\noalign{\smallskip}\hline
\end{tabular}
\end{table*}

\end{document}